\documentclass[twocolumn,notitlepage,prb,color,superscriptaddress,psfig,showpacs,amsmath,amssymb,nobibnotes,longbibliography]{revtex4-2}

\usepackage[colorlinks=true,citecolor=blue]{hyperref}%

\usepackage{epsf}
\usepackage{graphicx}
\usepackage{amssymb}
\usepackage{color}
\usepackage{float}
\usepackage{epstopdf}
\usepackage{natbib}
\usepackage{amsmath}
\usepackage{mathrsfs}

\usepackage{xspace}

\usepackage{soul}
\setstcolor{red}

\def\NPS{Ni$_{\text{2}}$P$_{\text{2}}$S$_{\text{6}}$}

\definecolor{darkgreen}{rgb}{0, 0.4, 0}

\begin{document}

\title{Low-energy excitations and magnetic anisotropy of the layered  \\ van der Waals antiferromagnet Ni$_{\text{2}}$P$_{\text{2}}$S$_{\text{6}}$}

\author{K.~Mehlawat}
\affiliation{Leibniz IFW Dresden, D-01069 Dresden, Germany}
\affiliation{Institute for Solid State and Materials Physics and W{\"u}rzburg-Dresden Cluster of Excellence ct.qmat, TU Dresden, D-01062 Dresden, Germany}
\author{A.~Alfonsov}
\affiliation{Leibniz IFW Dresden, D-01069 Dresden, Germany}
\author{S.~Selter}
\affiliation{Leibniz IFW Dresden, D-01069 Dresden, Germany}
\author{Y.~Shemerliuk}
\affiliation{Leibniz IFW Dresden, D-01069 Dresden, Germany}
\author{S.~Aswartham}
\affiliation{Leibniz IFW Dresden, D-01069 Dresden, Germany}
\author{B.~B\"uchner}
\affiliation{Leibniz IFW Dresden, D-01069 Dresden, Germany}
\affiliation{Institute for Solid State and Materials Physics and W{\"u}rzburg-Dresden Cluster of Excellence ct.qmat, TU Dresden, D-01062 Dresden, Germany}
\author{V.~Kataev}
\affiliation{Leibniz IFW Dresden, D-01069 Dresden, Germany}

\date{22 June 2022}
	
\begin{abstract}

The quasi-two-dimensional antiferromagnet \NPS\ belongs to the family of magnetic van der Waals compounds that provides a rich material base for the realization of fundamental models of quantum magnetism in low dimensions.  
Here, we report high-frequency/high-magnetic field electron spin resonance measurements on single crystals of \NPS. The results enable to reliably determine the positive, "easy"-plane type of the single ion anisotropy of the Ni$^{2+}$ ions with the upper limit of its magnitude $\lesssim 0.7$\,meV. The resonance response reveals strongly anisotropic spin fluctuations setting in shortly above the N\'eel temperature $T_{\rm N} = 158$\,K and extending in  the antiferromagnetically ordered state down to low temperatures. There, a low-energy magnon excitation gapped from the ground state by only $\sim 1$\,meV was found. 
This magnon  mode may explain unusual low temperature relaxation processes observed in the latest nuclear magnetic resonance experiments
and together with the estimate of the single ion anisotropy they should enable setting up a realistic spin model for an accurate description of magnetic properties of \NPS.

\end{abstract}
	
	\maketitle

\section{Introduction}

\begin{figure}[ht]
	\centering
	\includegraphics[clip,width=0.9\columnwidth]{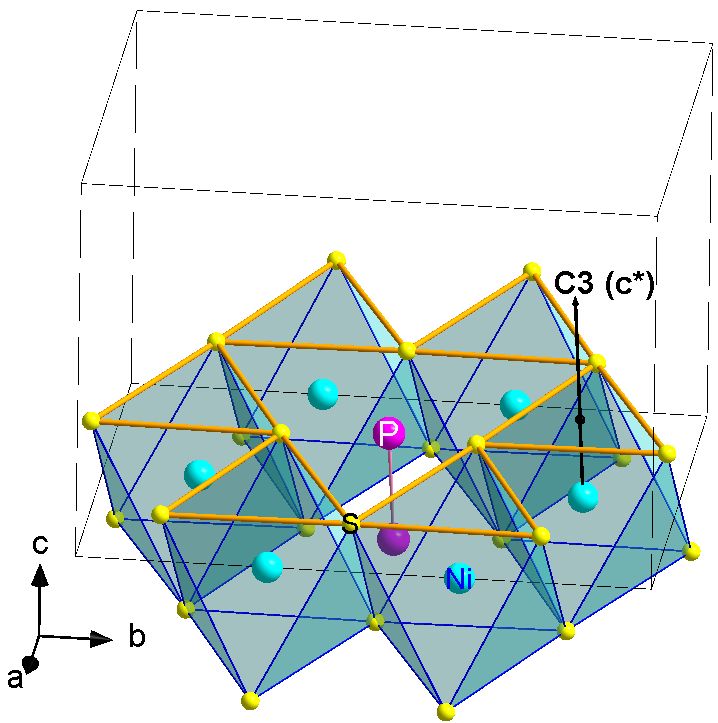}
	\caption{Structure of the individual \NPS\ layer: Ni$^{2+}$ ions (turquoise spheres) are octahedrally coordinated by six S ligands (yellow spheres). The NiS$_6$ octahedra make a honeycomb network in the $ab$~plane with the $C_3$ symmetry axis of the octahedra normal to the $ab$~plane ($c^*$~axis).  The P-P dumbbell occupies the void of each honeycomb. The honeycomb layers are stacked along the $c$~axis and form the crystallographic structure of the monoclinic symmetry, space group $C2/m$, with the $b$~axis along the somewhat longer Ni--Ni bond of the slightly distorted honeycomb and the $a$~axis normal to this bond  \cite{Klingen1973,Ouvrard1985}. } 
	\label{fig:structure}
\end{figure}

Magnetic van der Waals compounds have recently attracted significant attention both regarding new fundamental physical phenomena that they demonstrate and also in view of their potential for applications in next-generation spintronic devices \cite{Huang2017,Gong2017,Otrokov2019,Gong2019,Yang21}. One interesting subclass of these compounds is the family of transition metal (TM) tiophosphates $M_{\text{2}}$P$_{\text{2}}$S$_{\text{6}}$ where $M$ stands for a TM ion \cite{Brec86,Grasso02}. The spins associated with the $M$ ions are arranged in the crystallographic $ab$~plane on a two-dimensional (2D) honeycomb spin lattice (Fig.~\ref{fig:structure}). By virtue of the weak van der Waals coupling between the planes along the $c$~axis the honeycomb spin lattice often retains its quasi-2D character even in bulk crystals which thus provide an excellent platform for the realization and studies of the fundamental models of 2D magnets \cite{Pokrovsky90}. Depending on the choice of the $M$ ion different Hamiltonians can be realized in the $M_{\text{2}}$P$_{\text{2}}$S$_{\text{6}}$ family. In the case of $M$\,=\,Mn the system is a Heisenberg antiferromagnet \cite{Joy92,Wildes98} whereas the antiferromagnetism is of the Ising type for $M$\,=\,Fe \cite{Joy92,Lancon16,Selter21}. 

The situation is more complex in the case of \NPS. It orders antiferromagnetically (AFM) at a N\'eel temperature $T_{\rm N} = 155$\,K due to the residual interplane interactions and below $T_{\rm N}$ it demonstrates not only a significant easy-plane anisotropy but also an anisotropic magnetic response within the $ab$~plane enabling one to classify this material alternatively as anisotropic Heisenberg or the {\it XXZ} antiferromagnet \cite{Joy92,Wildes15,Lancon18,Selter21}. While in the above cited works the magnetic properties of \NPS\ were studied quite in detail by static magnetometry and also by elastic and inelastic neutron scattering (INS), addressing the spin dynamics by local spin probes received much less attention. Recently, a detailed nuclear magnetic resonance (NMR) study of single crystals of \NPS\ demonstrated the sensitivity of the $^{31}$P nuclear spin probe to quasi-2D correlations \cite{Dioguardi20}. However,  directly addressing the electronic spin system of \NPS\ with electron spin resonance (ESR) spectroscopy has not been attempted so far whereas this method was recently successfully applied for studies of 
%
%
several magnetic van der Waals compounds providing interesting insights onto the relationship between the spin dynamics and the electronic properties \cite{Otrokov2019,Zeisner2019,Khan2019,Saiz2019,Vidal2019,Zeisner2020,Saiz2021,Singamaneni2020,Ni2021,Alahmed2021,Sakurai2021,Alfonsov2021,Alfonsov2021b}.   

Here we present the results of a detailed high-frequency/high-field ESR (HF-ESR) spectroscopic study of well-characterized single crystalline samples of \NPS\ carried out in a broad range of excitation frequencies, magnetic fields and temperatures. Three temperature regimes could be identified. At $T>T_{\rm N}$ the HF-ESR parameters are almost independent of the direction of the applied field. The $g$~factors are slightly anisotropic characteristic of an easy-plane anisotropy of the Ni$^{2+}$ ions. Entering the AFM ordered state yielded a strongly anisotropic response indicating anisotropic spin fluctuations at the HF-ESR frequencies which persist at temperatures far below $T_{\rm N}$. Finally, at the base temperature of 4\,K  the frequency {\it versus} field dependence of the previously unobserved spin-wave mode was measured and its excitation energy was quantified. 
These findings shed light on  the previous INS, NMR and computational studies of \NPS\ and 
call for the development of the comprehensive theoretical model of spin excitations in \NPS.

\section{Experimental details}
\label{experimental}

High-quality single crystals of \NPS\ were grown by the chemical vapor transport technique with iodine as the transport agent. Details  of the growth and physical characterization of the crystals studied in this work are described in Ref.~\cite{Selter21}. 

HF-ESR measurements were carried out with two homemade setups. For measurements at frequencies in a range of 30--330\,GHz a vector network analyzer (PNA-X from Keysight Technologies) was employed for generation and detection of the microwave radiation. For measurements at higher frequencies up to 500\,GHz a modular amplifier/multiplier chain  (AMC from Virginia Diodes, Inc.) was used at the generation side in combination with  a hot electron InSb bolometer (QMC Instruments) at the detection side. Samples were placed into a probe head operational in the transmission mode in the Faraday configuration. The probe head was inserted into a $^4$He variable temperature unit of a 16\,T superconducting magnet system (Oxford Instruments). For angular dependent measurements a piezoelectric step-motor-based sample holder was mounted inside the probe head \cite{Fuchs17}. The ESR spectra were recorded in the field-sweep mode at selected constant microwave frequencies $\nu$ by continuously sweeping the field $\mu_0 H$ from 0 to 16\,T and then back to 0\,T. To obtain the linewidth $\Delta H$ and the resonance field $H_{\rm res}$ the ESR signals were fitted to the function 
\begin{eqnarray}
	\label{eq:fitfunction}
	f(H) \propto 1/(1+ x^2) + d\cdot x(1+x^2)\, ,\\ 
	x = (H - H_{\rm res})/(0.5\Delta H)\, . \nonumber
\end{eqnarray}
	The first and the second term in (\ref{eq:fitfunction}) are Lorentzian absorption and dispersion line shapes, respectively, and $d$ is the admixing coefficient. The latter term accounts for a possible admixture of the dispersive component to the measured signal which may arise under certain instrumental conditions in an HF-ESR experiment \cite{Reijerse2010}.

\section{Experimental results}

\begin{figure}[ht]
	\centering
	\includegraphics[clip,width=\columnwidth]{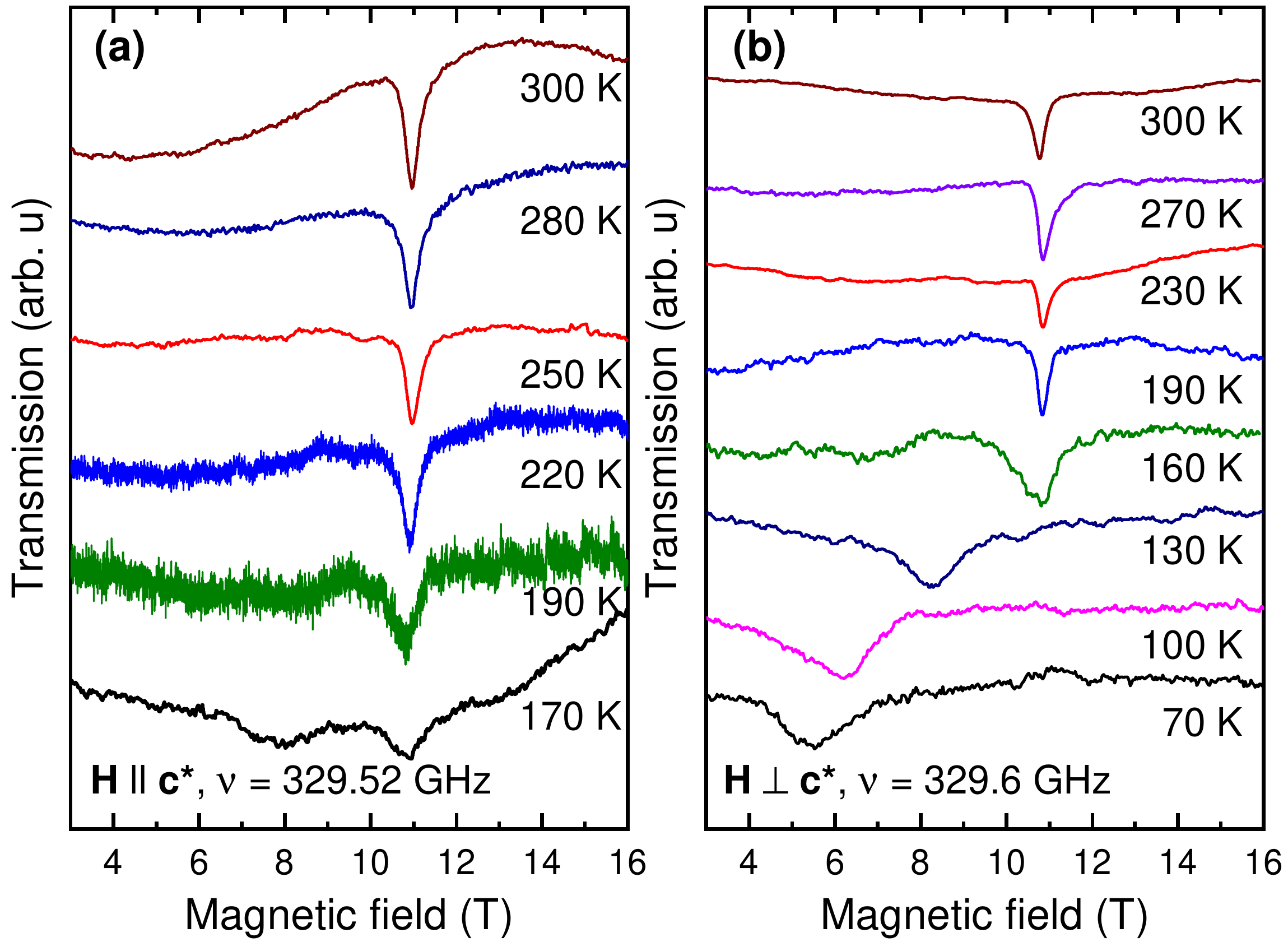}
	\caption{Temperature dependence of the HF-ESR signal at a fixed excitation frequency $\nu\approx 329$\,GHz for ${\bf H}\parallel {\bf c^\ast}$ (a) and ${\bf H}\perp {\bf c^\ast}$ (b).    } 
	\label{fig:spectra_paramag}
\end{figure}

\begin{figure}[ht]
	\centering
	\includegraphics[clip,width=\columnwidth]{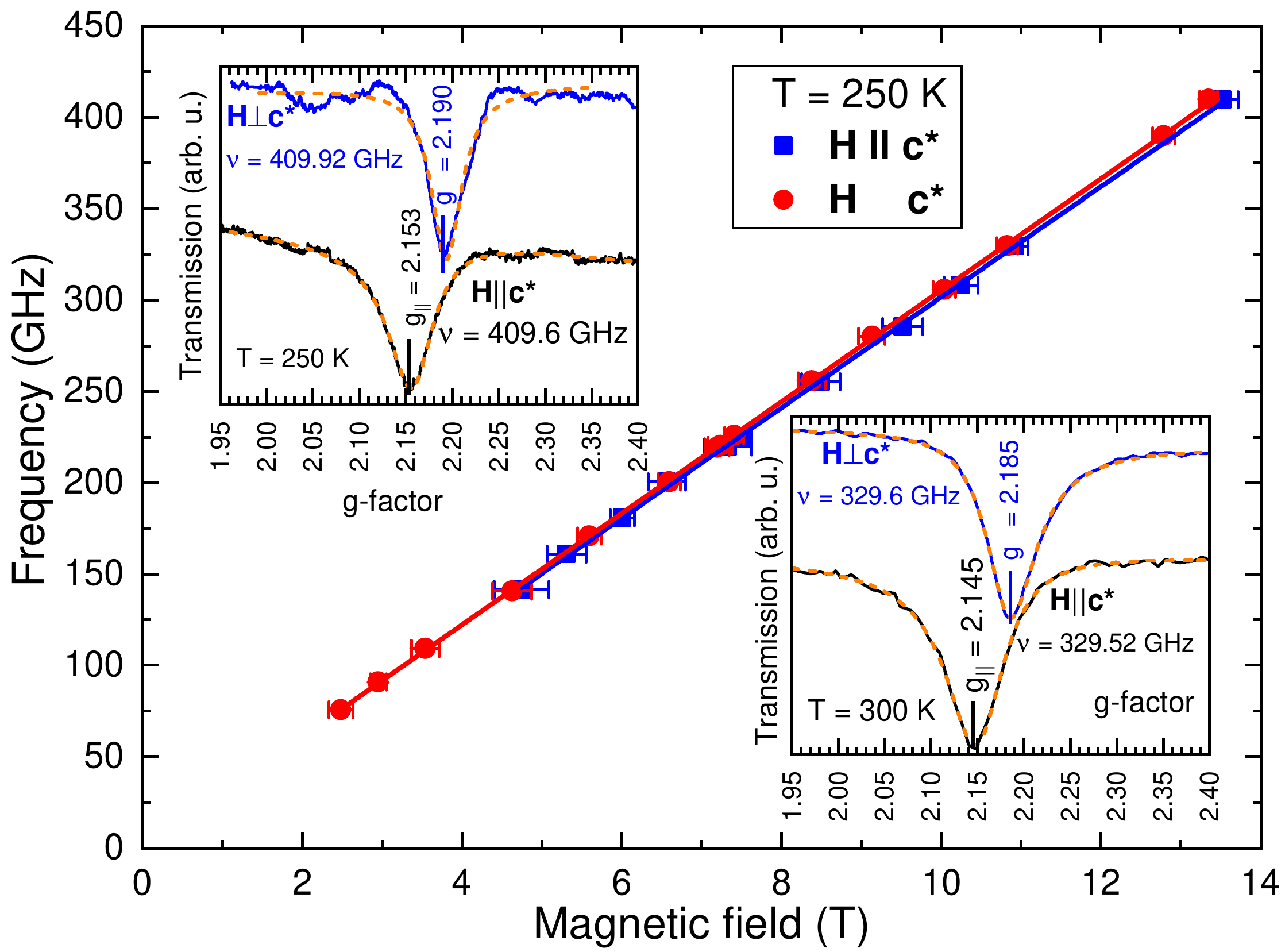}
	\caption{$\nu$ {\it versus} $H_{\rm res}$ dependence at $T = 250$\,K for ${\bf H}\parallel {\bf c^\ast}$ (blue squares) and ${\bf H}\perp {\bf c^\ast}$ (red circles). Solid lines represent the fit according to the resonance condition $h\nu = g\mu_{\rm B}\mu_0H_{\rm res}$. Insets: HF-ESR signals (black and blue solid lines) for ${\bf H}\parallel {\bf c^\ast}$ and ${\bf H}\perp {\bf c^\ast}$ plotted as a function of the $g$~factor $ g = h\nu/\mu_{\rm B}\mu_0H$  at $\nu \approx 410$\,GHz and $T = 300$\,K (top left) and $\nu \approx 330$\,GHz and $T = 250$\,K (bottom right). Orange dashed lines are fits to Eq.~(\ref{eq:fitfunction}). The $g$~values of the resonance peaks are indicated in the respective plots.   } 
	\label{fig:g-factors}
\end{figure}

Typical HF-ESR spectra of \NPS\ at various temperatures for the magnetic field ${\bf H}$ applied parallel and perpendicular to the $c^\ast$-axis are shown in Figs.~\ref{fig:spectra_paramag}(a) and \ref{fig:spectra_paramag}(b), respectively. Here, the $c^\ast$-axis is defined as the normal to the $ab$~planes (Fig.~\ref{fig:structure}). 
The frequency $\nu$ versus resonance field $H_{\rm res}$ dependence at $T = 250$\,K $> T_{\rm N} = 158$\,K \cite{Selter21} is presented in Fig.~\ref{fig:g-factors} for both field geometries. It follows the simple paramagnetic resonance condition $h\nu = g\mu_{\rm B}\mu_0H_{\rm res}$ yielding the slightly anisotropic $g$~factor values $g_{\parallel} = 2.16 \pm 0.02$ and $g_{\perp} = 2.18 \pm 0.02$ for ${\bf H}\parallel {\bf c^\ast}$ and ${\bf H}\perp {\bf c^\ast}$, respectively. Here $h$, $\mu_{\rm B}$ and $\mu_0$ are the Planck constant, Bohr magneton, and vacuum permeability, respectively. Since due to the difference of the $g$~factors  the spectral resolution increases with increasing the frequency (the resonance field) the accuracy of the determination of the $g$~values can be significantly improved by considering the spectra at high frequencies of 330 and 410\,GHz [Fig.~\ref{fig:g-factors} (insets)]. A shift of the signals to a larger $g$~factor by changing the orientation from ${\bf H}\parallel {\bf c^\ast}$ to ${\bf H}\perp {\bf c^\ast}$ is clearly visible. The observed shift corresponds to the $g$~values $g_{\parallel} = 2.149 \pm 0.004$ and $g_{\perp} = 2.188 \pm 0.004$.


\begin{figure}[ht]
	\centering
	\includegraphics[clip,width= \columnwidth]{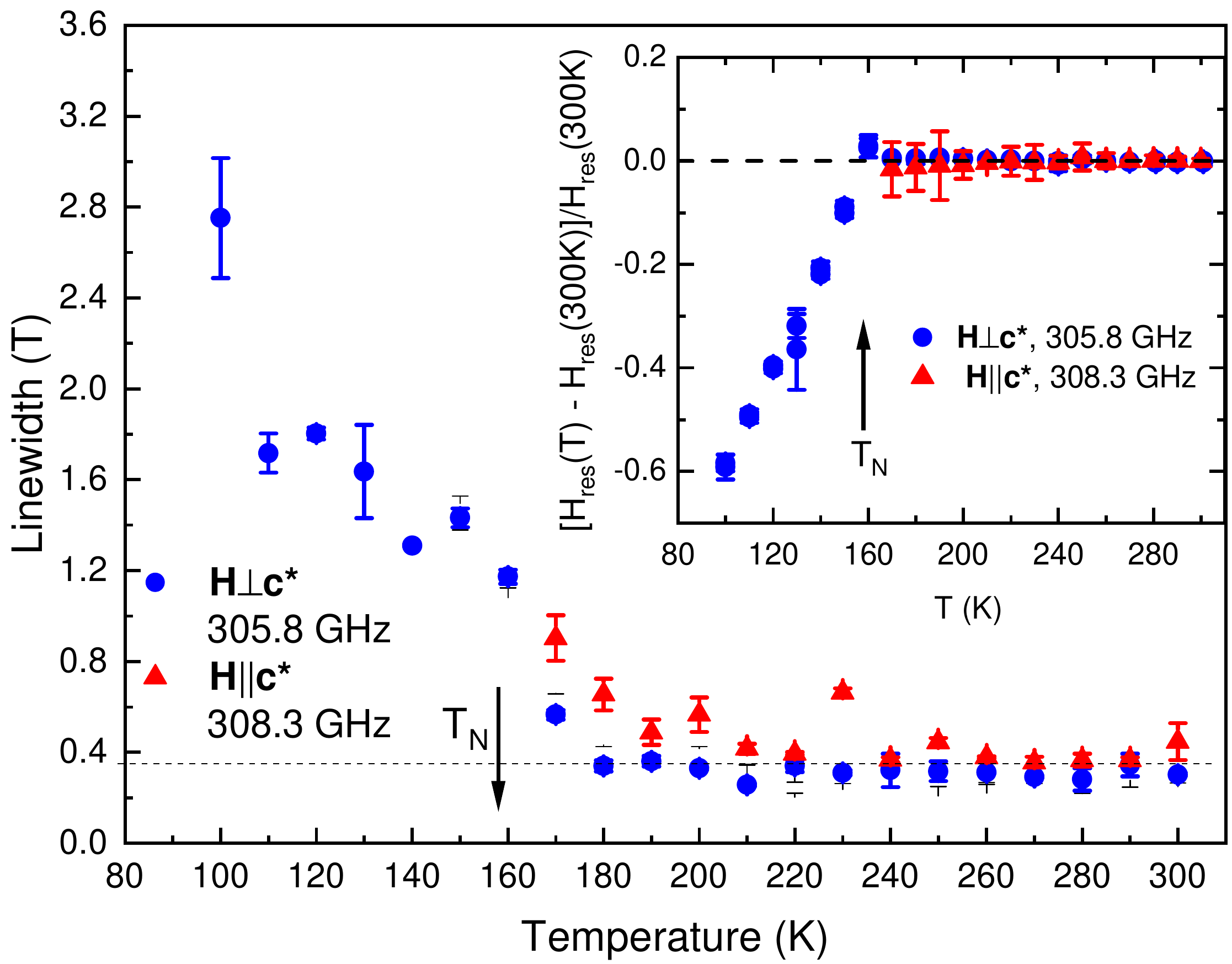}
	\caption{Temperature dependence of the linewidth (main panel) and of the relative shift (\ref{eq:relat_shift}) of the HF-ESR signal (inset) at two close frequencies of 308.3\,GHz for ${\bf H}\parallel {\bf c^\ast}$ (a) and of 305.8\,GHz  for ${\bf H}\perp {\bf c^\ast}$. } 
	\label{fig:linewidth_shift_T}
\end{figure}

%
The linewidth $\Delta H$ of the HF-ESR signal of \NPS\ in the paramagnetic state above $T_{\rm N}$ is practically isotropic (Fig.~\ref{fig:linewidth_shift_T}). 
The relative shift of the signal with respect to its position  at room temperature
\begin{equation}
\label{eq:relat_shift}
\delta H_{\rm res} = [H_{\rm res}(T) - H_{\rm res}(300\,K)]/ H_{\rm res}(300\,K)
\end{equation}	
%
%
is $T$~independent within the experimental uncertainty down to $T_{\rm N}$ and is nearly zero [Fig.~\ref{fig:linewidth_shift_T} (inset)], while $\Delta H$ begins to increase below 180\,K for both magnetic field geometries. Remarkably, at the vicinity of $T_{\rm N}$ the signal for ${\bf H}\parallel {\bf c^\ast}$ rapidly wipes out [Fig.~\ref{fig:spectra_paramag}(a)] and cannot be detected at lower temperatures in the available frequency range. In contrast, the signal for ${\bf H}\perp {\bf c^\ast}$ continuously broadens upon entering the magnetically ordered state and strongly shifts to lower fields [Fig.~\ref{fig:spectra_paramag}(b) and Fig.~\ref{fig:linewidth_shift_T}]. 

\begin{figure}[ht]
	\centering
	\includegraphics[clip,width= 0.9\columnwidth]{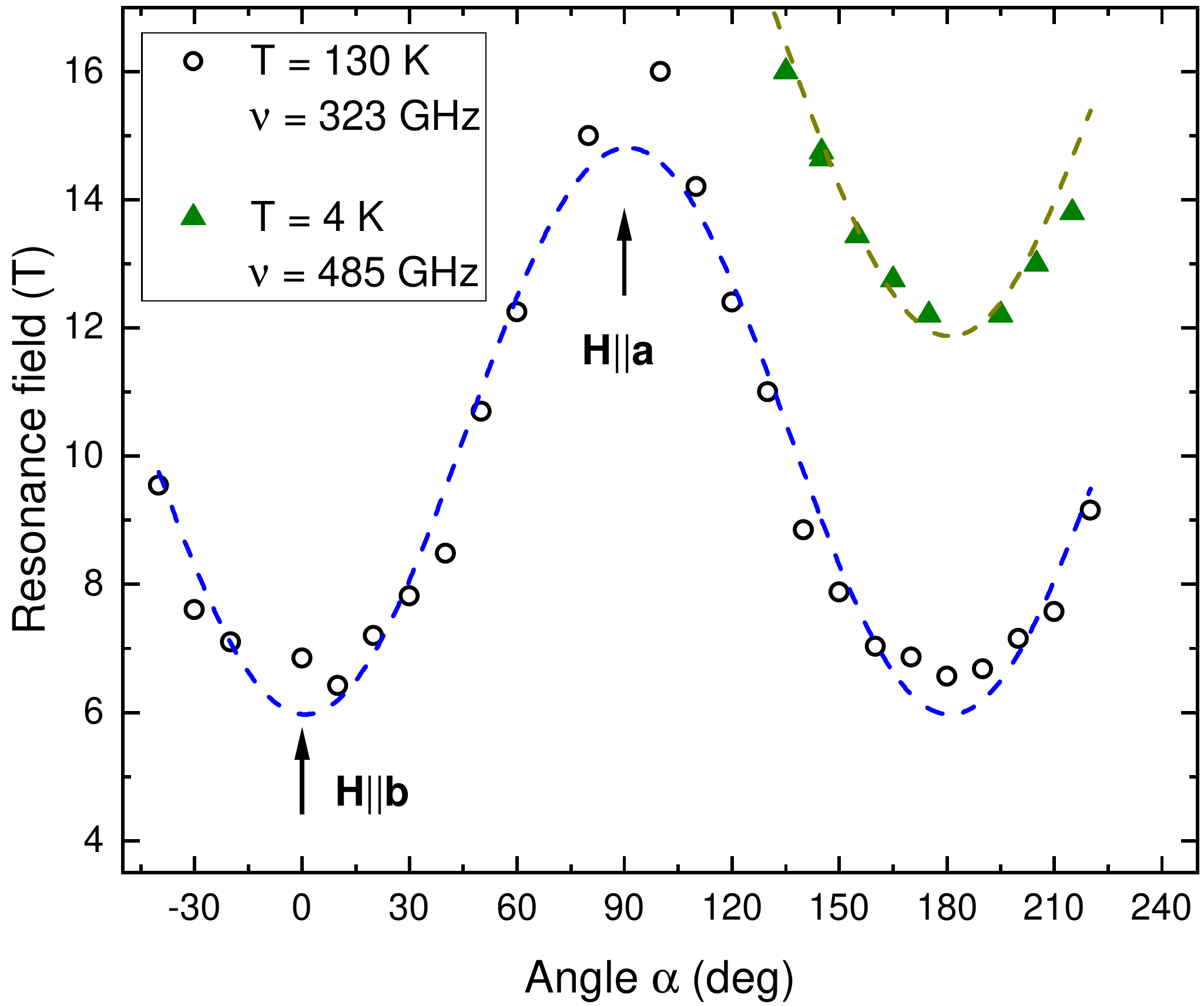}
	\caption{Dependence of the resonance field $H_{\rm res}$ on the angle $\alpha$ which the field ${\bf H}$ applied in the $ab$~plane makes with the $b$~axis at $T = 130$\,K and $\nu = 323$\,GHz (circles), and at $T = 4$\,K and $\nu = 485$\,GHz (triangles). Dashed lines are fits to the function $H_{\rm res}(\alpha) = A + B\sin^2(\alpha)$.  } 
	\label{fig:angular_dep}
\end{figure}

Measurements of the HF-ESR signal at $T < T_{\rm N}$ and at different orientations of the applied magnetic field within the $ab$~plane revealed a significant angular dependence of its position with the 180$^\circ$ periodicity which follows a simple sine law $H_{\rm res}(\alpha) = A + B\sin^2(\alpha)$ (Fig.~\ref{fig:angular_dep}). Here, $\alpha$ is the angle between ${\bf H}$ and the $b$~axis. The anisotropy of the resonance field with $H_{\rm res}(\alpha = 0) < H_{\rm res}(\alpha = 90^\circ)$ is in agreement with the anisotropy of the static magnetization $M_{\rm b} > M_{\rm a}$ \cite{Wildes15,Selter21,Mperiodicity} since for a stronger magnetized direction less external field is needed to reach the resonance condition. 
According to the neutron diffraction data in Ref.~\cite{Wildes15} the spins in the AFM ordered state of \NPS\ form in the $ab$~plane zigzag ferromagnetic chains along the $a$~axis AFM coupled along the orthogonal $b$~axis. Thus, the chain direction is the magnetic "easy" axis with the  magnetization smaller than that for the "hard" $b$~axis. Formation of the threefold  magnetic domain structure which could be possibly present in a slightly distorted honeycomb lattice was not reported in Ref.~\cite{Wildes15} and is not evident in our data.
The dependence $H_{\rm res}(\alpha)$ at $T = 4$\,K, shown in Fig.~\ref{fig:angular_dep} for comparison, could be obtained only in the limited range of angles since the signal at this low temperature could be measured only at much higher frequencies due to the opening of the AFM excitation gap (see below) and thus for many orientations the resonance field was out of the available field range.

\begin{figure}[ht]
	\centering
	\includegraphics[clip,width= \columnwidth]{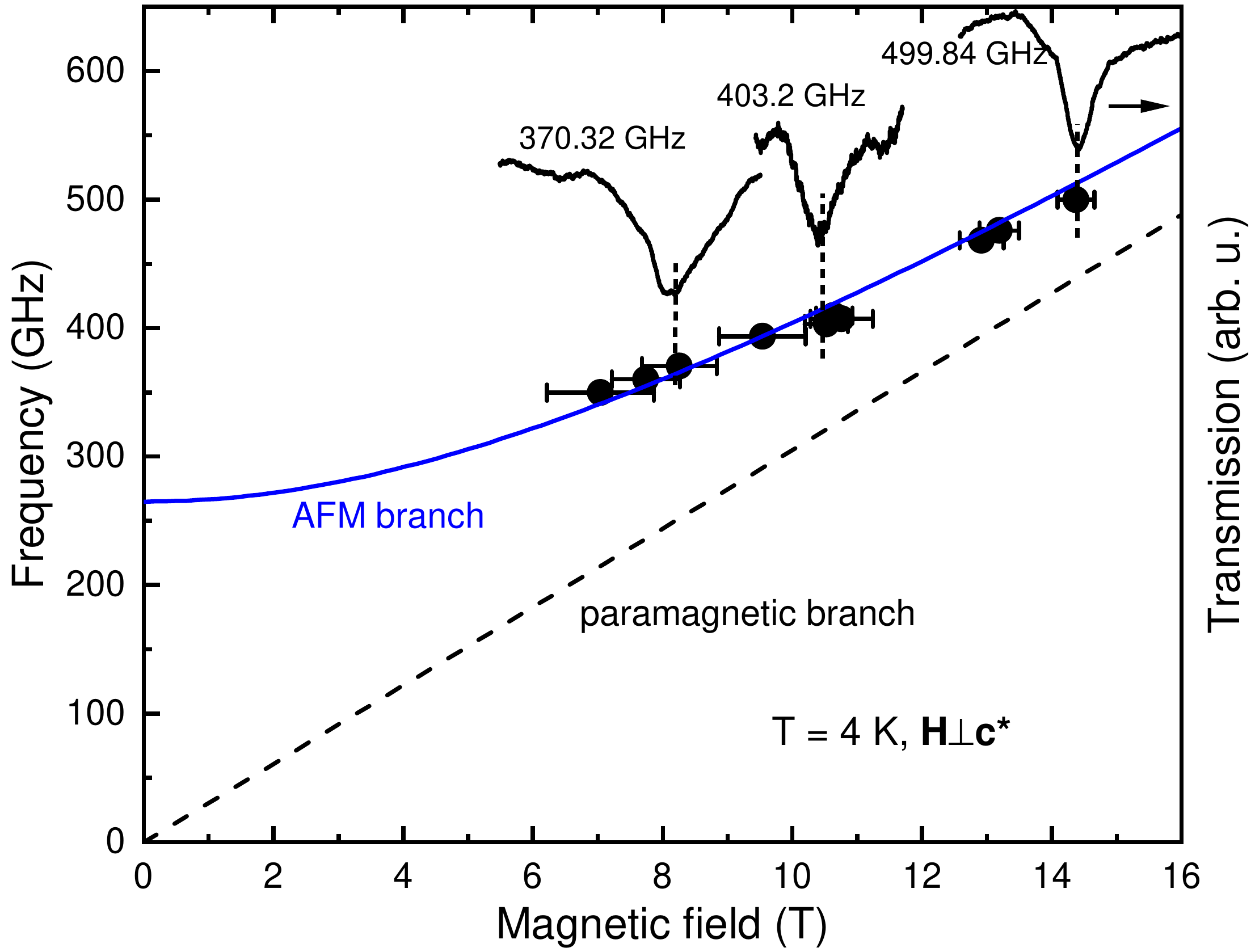}
	\caption{Left vertical scale: $\nu$ {\it versus} $H_{\rm res}$ dependence of the HF-ESR signal at $T = 4$\,K for ${\bf H}\perp {\bf c^\ast}$ (circles). Solid line depicts the fit of the data to Eq.~(\ref{eq:AFM_branch}).
	Dashed line corresponds to the paramagnetic resonance condition $\nu = h^{-1}g_{\perp}\mu_{\rm B}\mu_0H_{\rm res}$. Right vertical scale: HF-ESR signals at selected frequencies.
		} 
	\label{fig:angular_AFM_branch_4K}
\end{figure}

Below $T\sim 70$\,K the HF-ESR signal for ${\bf H}\perp {\bf c^\ast}$ is not observable anymore in the available frequency range presumably due to its significant broadening. However, the signal recovers at the base temperature of 4\,K where its $\nu(H_{\rm res})$ dependence, referred to hereafter as the AFM branch, can be measured with confidence (Fig.~\ref{fig:angular_AFM_branch_4K}).  
In contrast to the paramagnetic state where the position of the signal follows the linear resonance condition $h\nu = g\mu_{\rm B}\mu_0H_{\rm res}$ [Fig.~\ref{fig:spectra_paramag}(c)],  the AFM branch is shifted upwards significantly and gets apparently nonlinear. The "flattening" of this branch with lowering the frequency is reflected in the broadening of the signal because in the employed HF-ESR setup the spectra are recorded at a given fixed frequency by sweeping the magnetic field. Finally, the signal cannot be detected at $\nu < 350$\,GHz suggesting the presence of an energy gap for the AFM excitations.

\section{Discussion}

\subsection{Paramagnetic state}

The obtained $g$~values $g_{\parallel} = 2.149 \pm 0.004$ and $g_{\perp} = 2.188 \pm 0.004$ (Fig.~\ref{fig:g-factors}) fall into the common range of the $g$~factors found for the Ni$^{2+}$ (3$d^6$, $S = 1$) ions in the octahedral ligand coordination \cite{AbragamBleaney}. The observed anisotropy with $g_\perp > g_\parallel$ indicates a splitting of the $S = 1$ spin triplet into a singlet $\mid0\rangle$ and a doublet $\mid\pm 1\rangle$ due to the second-order spin-orbit coupling effect in the presence of the trigonal distortion of the ligand octahedra along the $c^*$~axis which can be parameterized using Hamiltonian \cite{AbragamBleaney} 
\begin{equation}
\label{eq:hamilton}
\mathcal{H}=D[S_z^2+(1/3)S(S+1)]\, .
\end{equation}
%
%
Here $D$ is the single-ion anisotropy (SIA) parameter.  
In the case of isolated Ni$^{2+}$ ions the initial (zero-field) splitting of the spin levels due to a finite value of $D$ should yield two ESR peaks corresponding to the transitions $|+1\rangle\!\!\leftrightarrows | 0\rangle$ and $|-1\rangle\!\!\leftrightarrows |0\rangle$ separated in field by $2| D|/g\mu_{\rm B}\mu_0$ \cite{AbragamBleaney}. Here $D = E_{|\pm 1\rangle} - E_{|0\rangle} $ is the energy difference of the zero-field split levels. However, for the interacting paramagnetic Ni centers  in concentrated magnets such as \NPS, the two signals merge into a single line at the midresonance field due to the exchange narrowing effect \cite{Smirnov2008}. This complication prevents a direct quantification of the magnitude of $D$ from the distance between the two fine-structure-split  ESR lines. Nevertheless, the sign and an order of magnitude of the SIA in \NPS\ can be estimated considering the relation between $D$ and $g$~factors in the form $D = \lambda (g_{\parallel} - g_{\perp})/2$ \cite{AbragamBleaney}. Here $\lambda$ is the spin-orbit coupling constant. Taking the Ni$^{2+}$ free ion value of $\lambda \sim -300$\,cm$^{-1} = -87$\,meV \cite{Ballhausen1962} and the experimentally determined $g$~factors one obtains $D\sim 0.7$\,meV. One should note, however, that the magnitude of $D$ might be overestimated since in a covalent solid such as \NPS\ $|\lambda|$ could be smaller than in the ionic limit.

The positive sign of $D$ implies that the $c^*$~axis should be the hard axis for the Ni spins in agreement with the density functional theory (DFT) result~\cite{Olsen2021}. Multireference configuration-interaction calculations are even in the quantitative agreement with the HF-ESR estimate \cite{Dioguardi20}. $D >0$ is compatible with the experimentally determined almost in-plane zigzag spin structure in the AFM ordered state of \NPS\ with the spins aligned along the $a$~axis~\cite{Wildes15}. This secondary, in-plane easy axis may possibly arise due to the spin-orbit coupling (SOC) driven anisotropy of the Heisenberg superexchange in a combination with dipole-dipole interactions \cite{Kim2021}. Indeed, such an easy-plane structure is found to be most stable in the DFT\,+\,$U$\,+\,SOC calculations~\cite{Koo2021}. Here, $U$ is the on-site electron repulsion energy. The positive $D$ would also naturally explain the suppression of the AFM order in the monolayer of \NPS\ \cite{Kim2019} because according to the Mermin-Wagner theorem a magnetic out-of-plane easy axis is required to stabilize magnetic order in a 2D spin lattice at $T>0$ \cite{commentKim2021}. At odds with these findings, the easy-axis type of SIA was derived from the analysis of the INS data~\cite{Lancon18}.

It is remarkable that both the resonance field $H_{\rm res}$ and the linewidth $\Delta H$ of the HF-ESR signal in \NPS\ remain nearly  isotropic and in particular $T$~independent by cooling the sample down to temperatures close to $T_{\rm N}$ (Fig.~\ref{fig:linewidth_shift_T}). Usually, in the quasi-2D spin systems the ESR line broadening and shift occur at $T> T_{\rm N}$ due to the growth of the in-plane spin-spin correlations resulting in a development of slowly fluctuating short-range order \cite{Benner1990}. Specifically, a distribution of the local fields and shortening of the spin-spin relaxation time due to the slowing down of the spin fluctuations increase the ESR linewidth. 

 The onset of the short-range ordered regime is typically associated with the broad maximum in the static magnetic susceptibility $\chi(T)$ which occurs in \NPS\ at $T_{\chi_{\rm max}} \sim 260$\,K \cite{Selter21}. It manifested in an anomalous  breakdown of the scaling between the $^{31}$P NMR shift and static $\chi$ but, interestingly, the NMR linewidth and the relaxation rate $1/T_1$ remained practically unchanged down to $T_{\rm N}$ \cite{Dioguardi20}. This suggests that approaching $T_{\rm N}$ from above the electron spin dynamics in \NPS\ still remains much faster than the NMR time window of the order of 10--100\,$\mu$s. On a much shorter, nano- to picosecond time scale of HF-ESR the slowing down of the spin dynamics becomes visible in the increase of $\Delta H$ only below 180\,K, i.e., just at $T \lesssim 1.1T_{\rm N}$. Such a short temperature interval above $T_{\rm N}$ for the manifestation of the low-D dynamic spin correlations suggests that from the viewpoint of local spin probes, such as NMR and particularly HF-ESR, \NPS\ can be considered rather as a quasi-3D than a quasi-2D spin system, i.e., the interlayer coupling should be sufficiently strong.   Indeed, a relatively high AFM ordering temperature is maintained only in bulk single crystals and is suppressed towards the pure 2D limit, which implies the significance of the coupling in the third dimension for the stabilization of magnetic order in a quasi-3D system with the easy-plane magnetic anisotropy~\cite{Kim2019}.

\subsection{Transition to the AFM ordered state}

For ${\bf H}\perp {\bf c^\ast}$,  the shift of the HF-ESR signal at $T<T_{\rm N}$ from its paramagnetic position can be ascribed to the opening of the energy gap for spin excitations arising due to magnetic anisotropy \cite{Turov}. Concomitantly, $\Delta H$ continue to increase strongly without tendency to saturate, which indicates significant spin fluctuations also in the AFM long-range ordered state of \NPS. In this field geometry both in-plane and out-of-plane fluctuations should contribute to the line broadening, while for  ${\bf H}\parallel {\bf c^\ast}$ the in-plane fluctuations make a dominant contribution (see, e.g., the Supplemental Material in Ref.~\cite{Alfonsov2021} for details). One can conjecture that a wipeout of the signal for the out-of-plane orientation of ${\bf H}$ at $T<T_{\rm N}$  might be a consequence of the sharp boosting of the in-plane spin fluctuations upon establishment of the long-range magnetic order. Eventually below $T\sim 70$\,K the signal gets unobservable  also for the other field geometry indicating a strong spectral density of the spin fluctuations at HF-ESR frequencies persisting in the AFM ordered state of \NPS\ far below~$T_{\rm N}$.

\subsection{Magnon gap}

Eventually spin fluctuations apparently cease at a low temperature of 4\,K enabling one to detect the HF-ESR signal in the ${\bf H}\perp {\bf c^\ast}$ configuration. Considering its specific frequency versus field dependence (Fig.~\ref{fig:angular_AFM_branch_4K}) one can ascribe this resonance mode with confidence  to the lowest in energy uniform ($q = 0$ wave vector) spin wave (magnon) excitation in \NPS\  (note that owing to the long wavelength of the applied microwave radiation it is usually not possible in an ESR experiment to excite the modes with a finite momentum transfer). 

The data can be reasonably well fitted with the function describing the AFM branch for the hard direction of a two-sublattice collinear antiferromagnet \cite{Turov} 
\begin{equation}
\label{eq:AFM_branch}
	\nu =h^{-1}[(g_{\perp}\mu_{\rm B}\mu_0H_{\rm res})^2+\Delta^2]^{1/2}
\end{equation}
%
%
with the in-plane $g$~factor $g_{\perp} \simeq  2.19$ obtained from the measurements in the paramagnetic state (Fig.~\ref{fig:g-factors}). The fit yields the excitation gap $\Delta = 260$\,GHz (1.07\,meV). The choice of the fit function appears reasonable since the $\nu(H_{\rm res})$ dependence was measured for the direction of the applied field close to the $b$~axis which is the hard in-plane direction of the AFM zigzag spin structure of \NPS\ \cite{Lancon18}.

Observation of such a low-energy mode with an excitation gap in the zero magnetic field limit amounting to only $\Delta \approx 1$\,meV sheds light into the spectrum of magnetic excitations in \NPS. The energy spectrum of the spin waves was studied recently by INS on single crystals of \NPS\  in Ref.~\cite{Lancon18}. It was concluded from the analysis of the magnon dispersions that the minimum excitation gap lies at the center of the 2D Brillouin zone ($\Gamma$-point) and amounts to $E_{\Gamma}\sim 7$\,meV. It followed from the same analysis that the "competing" gap of a similar magnitude $E_{C}$ should be present at the Brillouin zone corner ($C$-point). It was argued that the closeness of the magnitudes of these two gaps could give rise to the instability of the magnetic structure of \NPS. It appears now that the low-energy mode with the magnitude $\Delta \ll E_{\Gamma}, E_{C}$ found by HF-ESR was overlooked in the INS experiment possibly due to the problems with the energy resolution at small scattering vectors $q \approx 0$ where strong elastic scattering dominates the total response. Therefore, accounting for this mode in the excitation spectrum calls for the reanalysis of the exchange and anisotropy constants estimated from the INS data in Ref.~\cite{Lancon18}.

Furthermore, the occurrence of the low-energy magnon excitation with the gap $\Delta \approx 1$\,meV (12\,K) much smaller than $E_{\Gamma}, E_{C} \sim 7$\,meV (81\,K) may explain the puzzling observation of the power-law dependence of the $^{31}$P NMR relaxation rate $1/T_1 \propto T^5$ down to temperatures significantly smaller than 80\,K instead of the expected gap-like-activated behavior \cite{Dioguardi20}. This kind of $\propto T^5$ dependence is typical for the relaxation via the three-magnon scattering process and holds at temperatures higher than the magnon gap \cite{Beeman1968}. The low-lying excitation observed by HF-ESR is thus likely to provide an additional relaxation channel for nuclear spins at $T< E_{\Gamma}, E_{C} \sim 80$\,K.       

A possible instability of the spin structure in \NPS\ due to competing gaps at different points of the Brillouin zone that -- as suggested in Ref.~\cite{Lancon18} --  can be coupled to strong phonons in the same energy range and may thus give rise to enhanced fluctuations below $T_{\rm N}$. This provides a potential explanation of a continuous broadening of the HF-ESR signal and its disappearance in the AFM ordered state. Only when phonons  "freeze-out" at a significantly low temperature does the signal albeit broadened recover again.

\section{Conclusions}

In summary, we have performed a detailed HF-ESR spectroscopic study of the single-crystalline samples of the van der Waals compound \NPS, a member of the family of TM tiophosphates hosting  
by the virtue of the layered crystal structure a stack of 2D honeycomb spin planes weakly coupled in the third dimension. From the analysis of the $g$~tensor in the paramagnetic state above the AFM ordering temperature $T_{\rm N} = 158$\,K  we determined the positive sign of the SIA constant $D$ of the Ni$^{2+}$ ions and estimated its upper limit to be $D \lesssim 0.7$\,meV. 
This result supports computational predictions of an easy-plane type of  SIA and of the energetic stability of the in-plane order of Ni spins in \NPS\ \cite{Dioguardi20,Olsen2021,Koo2021}, in agreement with the experimentally determined spin structure~\cite{Wildes15} but at variance with the conclusion of the INS work  in Ref.~\cite{Lancon18} proposing the easy-axis type of SIA. 

A critical broadening of the HF-ESR signal usually observed in quasi-2D magnets far above $T_{\rm N}$ due to the development of the slowly fluctuating 2D short-range order is found in \NPS\ only in the vicinity of $T_{\rm N}$. This suggests a significant interlayer coupling which could be responsible for the high 3D AFM ordering temperature of this compound. 

A nearly isotropic HF-ESR response in the paramagnetic state of \NPS\  gets strongly anisotropic in the ordered state indicating strong and anisotropic spin fluctuations at HF-ESR frequencies. 
The fluctuations cease at low temperatures, enabling one to measure the $\nu(H_{\rm res})$ dependence of the in-plane magnon excitation branch which has a small energy gap $\Delta \approx 1$\,meV in the zero field limit. The occurrence of this low-energy spin wave excitation not observed in the INS study \cite{Lancon18} explains the unexpected three-magnon-assisted $^{31}$P NMR relaxation process at low temperatures \cite{Dioguardi20}. Altogether our results call for the revisiting of the analysis of the spin wave excitations in \NPS\ where a sizable interlayer magnetic coupling might be considered as well and should stimulate fundamental theoretical understanding of magnetic excitations also in other van der Waals TM tiophosphates.

\section{ACKNOWLEDGMENTS}

The authors would like to thank L.~T.~Corredor Bohorquez and L.~Hozoi for helpful discussions. 
This work was supported by the Deutsche Forschungsgemeinschaft (DFG) through Grants No. KA1694/12-1 and No. AS523/4-1, and within the Collaborative Research Center SFB 1143 ``Correlated Magnetism – From Frustration to Topology'' (Project-id No. 247310070) and the Dresden-Würzburg Cluster of Excellence (EXC 2147) `` ct.qmat - Complexity and Topology in Quantum Matter'' (Project-id No. 390858490). K.M. acknowledges the Hallwachs–Röntgen Postdoc Program of ct.qmat for financial support.

\bibliography{NiPS_ESR_rev2}

\end{document}